\documentclass[11pt,conference,letterpaper]{IEEEtran}
\usepackage{times,amsmath,epsfig}

\usepackage{amsthm}
\usepackage{amsfonts}
\usepackage{amssymb}
\usepackage{graphicx}
\usepackage{xspace} %Fixing newcommand spacing
\usepackage[ruled, vlined, linesnumbered]{algorithm2e}
\usepackage[caption = false]{subfig}
\graphicspath{{figures/}}

\usepackage{csvsimple}
\usepackage{booktabs}
\usepackage{siunitx}
\usepackage{stfloats}
\usepackage{verbatim}

\theoremstyle{plain}
\newtheorem{theorem}{Theorem}%[section]
\newtheorem{lemma}[theorem]{Lemma}

\theoremstyle{definition}
\newtheorem{definition}[theorem]{Definition}

\newtheorem{remark}[theorem]{Remark}
\newtheorem{observation}[theorem]{Observation}

\newcommand{\E}{\mathop{\rm E\/}}

\newcommand{\simjoin}{{\;\bowtie_\lambda\;}}
\newcommand{\simil}{\ensuremath{\textnormal{sim}}}
\newcommand{\cp}{\textsc{Chosen Path}\xspace}
\newcommand{\cpsj}{\textsc{CPSJoin}\xspace}
\newcommand{\mh}{\textsc{MinHash}\xspace}
\newcommand{\blsh}{\textsc{BayesLSH}\xspace}
\newcommand{\all}{\textsc{AllPairs}\xspace}
\newcommand{\pp}{\textsc{PPJoin}\xspace}

\newcommand{\Osymbol}{{O}}

\newcommand{\BOx}[1]{\Osymbol(#1)}

\title{Scalable and Robust Set Similarity Join}

\author{
{Tobias Christiani, Rasmus Pagh, Johan Sivertsen}%
\vspace{1.6mm}\\
\fontsize{10}{10}\selectfont\itshape
IT University of Copenhagen, Denmark \\
\fontsize{9}{9}\selectfont\ttfamily\upshape
\{tobc, pagh, jovt\}@itu.dk\\
}

\begin{document}
\maketitle
\begin{abstract} 
Set similarity join is a fundamental and well-studied database operator.
It is usually studied in the \emph{exact} setting where the goal is to compute all pairs of sets that exceed a given similarity threshold (measured e.g.~as Jaccard similarity).
But set similarity join is often used in settings where 100\% recall may not be important --- indeed, where the exact set similarity join is itself only an approximation of the desired result set.

We present a new randomized algorithm for set similarity join that can achieve any desired recall up to 100\%, 
and show theoretically and empirically that it significantly improves on existing methods.
The present state-of-the-art exact methods are based on prefix-filtering, the performance of which depends on the data set having many rare tokens.
Our method is robust against the absence of such structure in the data. 
At 90\% recall our algorithm is often more than an order of magnitude faster than state-of-the-art exact methods, depending on how well a data set lends itself to prefix filtering.
Our experiments on benchmark data sets also show that the method is several times faster than comparable approximate methods.
Our algorithm makes use of recent theoretical advances in high-dimensional sketching and indexing that we believe to be of wider relevance to the data engineering community.
\end{abstract}

\section{Introduction}
It is increasingly important for data processing and analysis systems to be able to work with data that is imprecise, incomplete, or noisy.
\emph{Similarity join} has emerged as a fundamental primitive in data cleaning and entity resolution over the last decade~\cite{augsten2013similarity,Chaudhuri_ICDE06, sarawagi2004efficient}.
In this paper we focus on \emph{set similarity join}:
Given collections $R$ and $S$ of sets the task is to compute 
$$ R \simjoin S = \{(x,y)\in R\times S \; | \; \simil(x,y)\geq \lambda\}$$
where $\simil(\cdot, \cdot)$ is a similarity measure and $\lambda$ is a threshold parameter.
We deal with sets $x,y \subseteq \{1,\dots,d\}$, where the number $d$ of distinct tokens can be naturally thought of as the dimensionality of the data.

Many measures of set similarity exist~\cite{choi2010survey}, but perhaps the most well-known such measure is the \emph{Jaccard similarity},
$$J(x,y) = |x\cap y|/|x\cup y| \enspace .$$
For example, the sets $x = \{$\texttt{IT, University, Copenhagen}$\}$ and $y = \{$\texttt{University, Copenhagen, Denmark}$\}$  have Jaccard similarity $J(x,y) = 1/2$ which could suggest that they both correspond to the same entity.
In the context of entity resolution we want to find a set~$T$ that contains $(x,y)\in R\times S$ if and only if $x$ and $y$ correspond to the same entity.
The quality of the result can be measured in terms of \emph{precision} $|(R \simjoin S) \cap T|/|T|$ and \emph{recall} $|(R \simjoin S) \cap T|/|R \simjoin S|$, both of which should be as high as possible.
We will be interested in methods that achieve 100\% precision, but that might not have 100\% recall.
We refer to methods with 100\% recall as exact, and others as approximate.

\subsection{Our Contributions}
We present a new approximate set similarity join algorithm: Chosen Path Similarity Join (\textsc{CPSJoin}).
We cover its theoretical underpinnings, and show experimentally that it achieves high recall with a substantial speedup compared to state-of-the-art exact techniques.
The key ideas behind \cpsj are:
\begin{itemize}
	\item A new recursive filtering technique inspired by the recently proposed {\textsc ChosenPath} index for set similarity search~\cite{christiani2017set}, adding new ideas to make the method parameter-free, near-linear space, and adaptive to a given data set.
	\item Apply efficient sketches for estimating set similarity~\cite{li2011theory} that take advantage of modern hardware. 
\end{itemize}

We compare \cpsj to the exact set similarity join algorithms in the comprehensive empirical evaluation of Mann et al.~\cite{Mann2016}, using the same data sets, and to other approximate set similarity join methods suggested in the literature.
We find that \cpsj outperforms other approximate methods and scales better than exact methods when the sets are relatively large (100 tokens or more) and the similarity threshold is low (e.g.~Jaccard similarity~0.5) where we see speedups of more than an order of magnitude at 90\% recall.
The finding that exact methods are faster in the case of high similarity thresholds, when the average set size is small, and when sets have many rare tokens, 
whereas approximate methods are faster in the case of low similarity thresholds and when sets are large, is consistent with theory and is further corroborated by experiments on synthetic datasets.

\subsection{Related Work}\label{sec:related}
For space reasons we present just a sample of the most related previous work, and refer to the book of Augsten and B{\"o}hlen~\cite{augsten2013similarity} for a survey of algorithms for exact similarity join in relational databases, covering set similarity joins as well as joins based on string similarity.

\subsubsection*{Exact Similarity Join}
Early work on similarity join focused on the important special case of detecting \emph{near-duplicates} with similarity close to~1, see e.g.~\cite{broder2000identifying,sarawagi2004efficient}.
A sequence of results starting with the seminal paper of Bayardo et al.~\cite{Bayardo_WWW07} studied the range of thresholds that could be handled.
Recently, Mann et al.~\cite{Mann2016} conducted a comprehensive study of 7 state-of-the-art algorithms for exact set similarity join for Jaccard similarity threshold $\lambda \in \{0.5,0.6,0.7,0.8,0.9\}$.
These algorithms all use the idea of \emph{prefix filtering}~\cite{Chaudhuri_ICDE06}, which generates a sequence of candidate pairs of sets that includes all pairs with similarity above the threshold.
The methods differ in how much additional filtering is carried out.
For example,~\cite{xiao2011efficient} applies additional \emph{length} and \emph{suffix} filters to prune the candidate pairs.

Prefix filtering uses an inverted index that for each element stores a list of the sets in the collection containing that element.
Given a set $x$, assume that we wish to find all sets $y$ such that~$|x \cup y| > t$.
A valid result set $y$ must be contained in at least one of the inverted lists associated with any subset of $|x| - t$ elements of $x$, or we would have $|x \cup y| \leq t$. 
In particular, to speed up the search, prefix filtering looks at the elements of $x$ that have the shortest inverted lists.

The main finding by Mann et al.~is that while more advanced filtering techniques do yield speedups on some data sets, an optimized version of the basic prefix filtering method (referred to as ``ALL'') is always competitive within a factor 2.16, and most often the fastest of the algorithms.
For this reason we will be comparing our results against ALL.

\subsubsection*{Locality-Sensitive Hashing}
Locality-sensitive hashing (LSH) is a theoretically well-founded randomized method for generating candidate pairs~\cite{Gionis99}.
A family of locality-sensitive hash functions is a distribution over functions with the property that the probability that similar points (or sets in our case) are more likely to have the same function value.
We know only of a few papers using LSH techniques to solve similarity join.
Cohen et al.~\cite{cohen2001finding} used LSH techniques for set similarity join in a knowledge discovery context before the advent of prefix filtering.
They sketch a way of choosing parameters suitable for a given data set, but we are not aware of existing implementations of this approach.
Chakrabarti et al.~\cite{chakrabarti2015bayesian} improved plain LSH with an adaptive similarity estimation technique, \emph{BayesLSH}, that reduces the cost of checking candidate pairs and typically improves upon an implementation of the basic prefix filtering method by $2$--$20\times$.
Our experiments include a comparison against both methods~\cite{cohen2001finding,chakrabarti2015bayesian}.
We refer to the survey paper~\cite{pagh2015large} for an overview of newer theoretical developments on LSH-based similarity joins, but point out that these developments have not matured sufficiently to yield practical improvements.

\subsubsection*{Distance Estimation}
Similar to BayesLSH~\cite{chakrabarti2015bayesian} we make use of algorithms for similarity \emph{estimation}, but in contrast to BayesLSH we use algorithms that make use of bit-level parallelism.
This approach works when there exists a way of picking a random hash function $h$ such that
\begin{equation}\label{eq:lsh-able}
\Pr[h(x)=h(y)] = \simil(x,y)
\end{equation}
for every choice of sets $x$ and $y$.
Broder et al.~\cite{Broder_NETWORK97} presented such a hash function for Jaccard similarity, now known as ``minhash'' or ``minwise hashing''.
In the context of distance estimation, 1-bit minwise hashing of Li and K{\"o}nig~\cite{li2011theory} maps minhash values to a compact sketch, often using just 1 or 2 machine words.
Still, this is sufficient information to be able to estimate the Jaccard similarity of two sets $x$ and $y$ just based on the Hamming distance of their sketches. 

%\begin{comment}
\subsubsection*{Locality-Sensitive Mappings}
Several recent theoretical advances in high-dimensional indexing~\cite{andoni2017optimal,christiani2017framework,christiani2017set} have used an approach that can be seen as a generalization of LSH.
We refer to this approach as locality-sensitive \emph{mappings} (also known as locality-sensitive \emph{filters} in certain settings).
The idea is to construct a function $F$, mapping a set $x$ into a set of machine words, such that:
\begin{itemize}
	\item If $\simil(x,y)\geq \lambda$ then $F(x)\cap F(y)$ is nonempty with some fixed probability~$\varphi > 0$.
	\item If $\simil(x,y) < \lambda$, then the expected intersection size $\E[|F(x)\cap F(y)|]$ is ``small''.
\end{itemize}
Here the exact meaning of ``small'' depends on the difference~\mbox{$\lambda - \simil(x,y)$}, 
but in a nutshell, if it is the case that almost all pairs have similarity significantly below $\lambda$ then we can expect \mbox{$|F(x)\cap F(y)| = 0$} for almost all pairs.
Performing the similarity join amounts to identifying all candidate pairs $x,y$ for which $F(x)\cap F(y) \ne \varnothing$ (for example by creating an inverted index), 
and computing the similarity of each candidate pair.
To our knowledge these indexing methods have not been tried out in practice, probably because they are rather complicated.
An exception is the recent paper~\cite{christiani2017set}, which is relatively simple, and indeed our join algorithm is inspired by the index described in that paper.
%\end{comment}
\section{Preliminaries}

The \textsc{CPSJoin} algorithm solves the $(\lambda,\varphi)$-similarity join problem with a probabilistic guarantee on recall, formalized as follows:
\begin{definition}
	\label{def:simjoin}
	An algorithm solves the $(\lambda,\varphi)$-similarity join problem with threshold $\lambda\in (0,1)$ and recall probability $\varphi \in (0,1)$ if for every $(x, y) \in S \bowtie_{\lambda} R$ the output $L \subseteq S \bowtie_{\lambda} R$ of the algorithm satisfies $\Pr[(x, y) \in L] \geq \varphi$. 
\end{definition}

%\begin{comment}
It is important to note that the probability is over the random choices made by the algorithm, and \emph{not} over a random choice of $(x,y)$.
This means that for any $(x,y)\in S \bowtie_{\lambda} R$ the probability that the pair is \emph{not} reported in $r$ independent repetitions of the algorithm is bounded by $(1-\varphi)^r$.
For example if $\varphi = 0.9$ it takes just $r=3$ repetitions to bound the recall to at least $99.9\%$.
%\end{comment}

\subsection{Similarity Measures}\label{sec:reduction}
Our algorithm can be used with a broad range of similarity measures through randomized \emph{embeddings}.
This allows it to be used with, for example, Jaccard and cosine similarity thresholds.

Embeddings map data from one space to another while approximately preserving distances, with accuracy that can be tuned.
In our case we are interested in embeddings that map data to sets of tokens.
We can transform any so-called \emph{LSHable} similarity measure $\simil$, where we can choose $h$ to make (\ref{eq:lsh-able}) hold, into a set similarity measure by the following randomized embedding:
For a parameter $t$ pick hash functions $h_1,\dots,h_t$ independently from a family satisfying~(\ref{eq:lsh-able}).
The embedding of $x$ is the following set of size~$t$:
$$ f(x) = \{ (i,h_i(x)) \; | \; i=1,\dots,t \} \enspace .$$
It follows from~(\ref{eq:lsh-able}) that the expected size of the intersection $f(x)\cap f(y)$ is $t \cdot \simil(x,y)$.
Furthermore, it follows from standard concentration inequalities that the size of the intersection will be close to the expectation with high probability.
%\[\Pr\left[\left|\frac{|f(x)\cap f(y)|}t-\simil(x,y)\right|\geq %\sqrt{\frac{6\ln{t}}{t}}\simil(x,y)\right]\leq2t^{-\simil(x,y)}\]
For our experiments with Jaccard similarity thresholds $\geq0.5$,  we found that $t=64$ gave sufficient precision for $>90\%$ recall.

In summary we can perform the similarity join $R\simjoin S$ for any LSHable similarity measure by creating two corresponding relations $R'=\{f(x) \; | \; x\in R\}$ and $S'=\{f(y) \; | \; y\in S\}$, 
and computing $R'\simjoin S'$ with respect to the similarity measure
\begin{equation}\label{eq:bb}
B(f(x),f(y)) = |f(x)\cap f(y)|/t \enspace .
\end{equation}
This measure is the special case of \emph{Braun-Blanquet} similarity where the sets are known to have size~$t$~\cite{choi2010survey}.
Our implementation will take advantage of the set size $t$ being fixed, though it is easy to extend to general Braun-Blanquet similarity.

The class of LSHable similarity measures is large, as discussed in~\cite{chierichetti2015lsh}.
If approximation errors are tolerable, even \emph{edit distance} can be handled by our algorithm~\cite{chakraborty2016streaming,zhang2017embedjoin}.

\subsection{Notation}

We are interested in sets $S$ where an element, $x\in S$ is a set with elements from some universe $[d]=\{1,2,3,\cdots,d\}$.
To avoid confusion we sometimes use ``record'' for $x\in S$ and ``token'' for the elements of $x$.
Throughout this paper we will think of a record $x$ both as a set of tokens from $[d]$, as well as a vector from $\{0,1\}^d$, where:
\[
x_i=
\begin{cases}
1\text { if } i\in x\\
0\text { if } i\notin x
\end{cases}
\]
It is clear that %by imposing some order on the universe of elements 
these representations are equivalent.
The set $\{1,4,5\}$ is equivalent to $(1,0,0,1,1,0,\cdots,0)$, $\{1,d\}$ is equivalent to $(1,0,\cdots,0,1)$, etc.

\section{Overview of approach}\label{sec:overview}
Our high-level approach is recursive and works as follows.
To compute $R\simjoin S$ we consider each $x\in R$ and either:
\begin{enumerate}
	\item Compare $x$ to each record in $S$ (referred to as ``brute forcing'' $x$), or
	\item create several subproblems $S_i \simjoin R_i$ with $x\in R_i \subseteq R$, $S_i \subseteq S$, and solve them recursively.
\end{enumerate}
The approach of~\cite{christiani2017set} corresponds to choosing option 2 until reaching a certain level $k$ of the recursion, where we finish the recursion by choosing option~1.
This makes sense for certain worst-case data sets, but we propose an improved parameter-free method that is better at adapting to the given data distribution.
In our method the decision on which option to choose depends on the size of $S$ and the average similarity of $x$ to the records of $S$.
We choose option~1 if $S$ has size below some (constant) threshold, or if the average Braun-Blanquet similarity of $x$ and $S$, $\tfrac{1}{|S|}\sum_{y\in S} B(x,y)$, is close to the threshold~$\lambda$.
In the former case it is cheap to finish the recursion.
In the latter case many records $y\in S$ will have $B(x,y)$ larger than or close to $\lambda$, so we do not expect to be able to produce output pairs with $x$ in sublinear time in $|S|$.

If neither of these pruning conditions apply we choose option~2 and include $x$ in recursive subproblems as described below.
But first we note that the decision of which option to use can be made efficiently for each $x$, since the average similarity of pairs from $R\times S$ can be computed from token frequencies in time $\BOx{t|R|+ t|S|}$.
Pseudocode for a self-join version of \cpsj is provided in Algorithm~\ref{alg:cpsjoin} and~\ref{alg:bruteforce}.
\subsection{Recursion} 
We would like to ensure that for each pair $(x,y)\in R\simjoin S$ the pair is computed in one of the recursive subproblems, i.e., that $(x,y)\in R_i\simjoin S_i$ for some $i$.
In particular, we want the expected number of subproblems containing $(x,y)$ to be at least~1, i.e.,
\begin{equation}\label{eq:expect}
	\E[|\{i \;|\; (x,y)\in R_i\simjoin S_i\}|] \geq 1.
	\end{equation}
	%
	%Let $R'$ and $S'$ be the subsets of $R$ and $S$ that do not satisfy any of the pruning conditions.
	To achieve (\ref{eq:expect}) for every pair $(x,y)\in R\simjoin S$ we proceed as follows: for each $i\in \{1,\dots,d\}$ we recurse with probability $1/(\lambda t)$ on the subproblem $R_i\simjoin S_i$ with sets
	\begin{align*}
	R_i &= \{ x\in R \mid i\in x \}\\
	S_i &= \{ y\in S \mid i\in y \}
	\end{align*}
	where $t$ denotes the size of records in $R$ and $S$.
	It is not hard to check that (\ref{eq:expect}) is satisfied for every pair $(x,y)$ with $B(x,y)\geq\lambda$.
	Of course, expecting one subproblem to contain $(x,y)$ does not \emph{directly} imply a good probability that $(x,y)$ is contained in at least one subproblem.
	But it turns out that we can use results from the theory of branching processes to show such a bound;
	details are provided in section~\ref{sec:algorithm}.
%\end{itemize}

%!TEX root = simfilter.tex
%--------------------------------------------------------
\section{Chosen Path Similarity Join} \label{sec:algorithm}
%--------------------------------------------------------
The \textsc{CPSJoin} algorithm solves the $(\lambda,\varphi)$-set similarity join~(Definition~\ref{def:simjoin}) for every choice of $\lambda \in (0,1)$ and with a guarantee on $\varphi$ that we will lower bound in the analysis. 

To simplify the exposition we focus on a self-join version where we are given a set $S$ of $n$ subsets of $[d]$ and we wish to report $L \subseteq S \simjoin S$.
Handling a general join $S \simjoin R$ follows the overview in section~\ref{sec:overview} and requires no new ideas: Essentially consider a self-join on $S\cup R$ but make sure to consider only pairs in $S\times R$ for output.
We also make the simplifying assumption that all sets in $S$ have a fixed size~$t$.
As argued in section~\ref{sec:reduction} the general case can be reduced to this one by embedding.

%--------------------------------------------------------
\subsection{Description}
%--------------------------------------------------------
The \text{CPSJoin} algorithm (see Algorithm \ref{alg:cpsjoin} for pseudocode) works by recursively splitting the data set on elements of~$[d]$ that are selected according to a random process, 
forming a recursion tree with $S$ at the root and subsets of $S$ that are non-increasing in size as we get further down the tree.
The randomized splitting has the property that the probability of a pair of sets $(x,y)$ being in a random subproblem is increasing as a function of $|x\cap y|$.

Before each recursive splitting step we run the \textsc{BruteForce} subprocedure (see Algorithm \ref{alg:bruteforce} for pseudocode) that identifies subproblems that are best solved by brute force.
It has two parts:

1. If $S$ is below some constant size, controlled by the parameter \texttt{limit}, we report $S\bowtie_{\lambda}S$ exactly using a simple loop with $O(|S|^2)$ distance computations (\textsc{BruteForcePairs}) and exit the recursion.
In our experiments we have set \texttt{limit} to $250$, with the precise choice seemingly not having a large effect as shown experimentally in Section~\ref{sec:parameters}. 

2. If $S$ is larger than \texttt{limit} the second part activates:
for every $x \in S$ we check whether the expected number of distance computations involving $x$ is going to decrease by continuing the recursion.
If this is not the case, we immediately compare $x$ against every point in $S$ (\textsc{BruteForcePoint}), reporting close pairs, and proceed by removing $x$ from $S$.
The \textsc{BruteForce} procedure is then run again on the reduced set. 

This procedure where we choose to handle some points by brute force crucially separates our algorithm from many other approximate similarity join methods in the literature that typically are LSH-based~\cite{PaghSIMJOIN2015, cohen2001finding}.
By efficiently being able to remove points at the ``right'' time, before they generate too many expensive comparisons further down the tree,
we are able to beat the performance of other approximate similarity join techniques in both theory and practice.
Another benefit of this approach is that it reduces the number of parameters compared to the usual LSH setting where the depth of the tree has to be selected by the user.

\begin{algorithm}
\DontPrintSemicolon
\emph{For $j \in [d]$ initialize $S_j \leftarrow \varnothing$.}\;
$S \leftarrow \textsc{BruteForce}(S, \lambda)$\;
$r \leftarrow \textsc{SeedHashFunction}()$\; %\tcp*[l]{Random $h \colon [d] \to [0,1]$}
\For{$x \in S$} {
	\For{$j \in x$} {

		\lIf{$r(j) < \frac{1}{\lambda |x|}$}{$S_{j} \leftarrow S_{j} \cup \{ x \} $}
	}
}
\lFor{$S_{j} \neq \varnothing$} {$\textsc{CPSJoin}(S_{j}, \lambda)$}
\caption{\textsc{CPSJoin}$(S, \lambda)$} \label{alg:cpsjoin}
\end{algorithm}

\begin{algorithm}
\SetKwData{Limit}{limit}
\SetKwArray{Count}{count}
\SetKwInOut{Global}{Global parameters}
\DontPrintSemicolon
\Global{\Limit $\geq 1$, $\varepsilon \geq 0$.}
\emph{Initialize empty map \Count{\,} with default value $0$.}\;
\If{$|S| \leq$ \Limit} {
	$\textsc{BruteForcePairs}(S, \lambda)$\;
	\Return $\varnothing$\;
}
\For{$x \in S$} {
	\For{$j \in x$} {
		\Count{j} $\leftarrow$ \Count{j} + 1\;
	}
}
\For{$x \in S$} {
	\If{$\frac{1}{|S| - 1}\sum_{j \in x}($\Count{j}$ - 1)/t > (1-\varepsilon)\lambda$} {
		$\textsc{BruteForcePoint}(S, x, \lambda)$\;
		\Return $\textsc{BruteForce}(S \setminus \{ x \}, \lambda)$\;
	} 
}
\Return $S$\;
	
\caption{\textsc{BruteForce}$(S, \lambda)$} \label{alg:bruteforce}
\end{algorithm}
%--------------------------------------------------------
\subsection{Comparison to Chosen Path}
% --------------------------------------------------------
\label{sec:comp-chos-path}
The \textsc{CPSJoin} algorithm is inspired by the \textsc{Chosen Path} algorithm~\cite{christiani2017set} for the approximate near neighbor problem 
and uses the same underlying random splitting tree that we will refer to as the Chosen Path Tree.
In the approximate near neighbor problem, the task is to construct a data structure that takes a query point and correctly reports an approximate near neighbor, if such a point exists in the data set.
Using the \textsc{Chosen Path} data structure directly to solve the $(\lambda,\varphi)$-set similarity join problem has several drawbacks that we avoid in the \textsc{CPSJoin} algorithm.
First, the \textsc{Chosen Path} data structure is parameterized in a non-adaptive way to provide guarantees for worst-case data, 
vastly increasing the amount of work done compared to the optimal parameterization when data is not worst-case.
Our recursion rule avoids this and instead continuously adapts to the distribution of distances as we traverse down the tree. 
Secondly, the data structure uses space $\BOx{n^{1+\rho}}$ where $\rho > 0$, storing the Chosen Path Tree of size $\BOx{n^\rho}$ for every data point. 
The \textsc{CPSJoin} algorithm, instead of storing the whole tree, essentially performs a depth-first traversal, using only near-linear space in $n$ in addition to the space required to store the output.
Finally, the \textsc{Chosen Path} data structure only has to report a single point that is approximately similar to a query point, and can report points with similarity $< \lambda$.
To solve the approximate similarity join problem the \textsc{CPSJoin} algorithm has to satisfy reporting guarantees for \emph{every} pair of points $(x, y)$ in the exact join.
%--------------------------------------------------------
\subsection{Analysis} \label{sec:analysis}
%--------------------------------------------------------
The Chosen Path Tree for a set $x \subseteq [d]$ is defined by a random process: 
at each node, starting from the root, we sample a random hash function $r \colon [d] \to [0,1]$ and construct children for every element $j \in x$ such that $r(j) < \frac{1}{\lambda |x|}$.
Nodes at depth $k$ in the tree are identified by their path $p = (j_1, \dots, j_k)$. 
Formally, the set of nodes at depth $k > 0$ in the Chosen Path Tree for $x$ is given by
\begin{equation*}
	F_{k}(x) = \left\{ p \circ j \mid p \in F_{k-1}(x) \land r_{p}(j) < \frac{x_j}{\lambda |x|} \right\}
\end{equation*}
where $p \circ j$ denotes vector concatenation and $F_{0}(x) = \varnothing$. % \{()\}$ is the set containing only the empty vector.
The subset of the data set $S$ that survives to a node with path $p = (j_1, \dots, j_{k})$ is given by
\begin{equation*}
	S_{p} = \{ x \in S \mid x_{j_1} = 1 \land \dots \land x_{j_{k}} = 1 \}.
\end{equation*}
The random process underlying the Chosen Path Tree belongs to the well studied class of Galton-Watson branching processes~\cite{harris2002theory}.
Originally these where devised to answer questions about the growth and decline of family names in a model of population growth assuming i.i.d.\ offspring for every member of the population across generations~\cite{watson1875}.
In order to make statements about the properties of the \textsc{CPSJoin} algorithm we study in turn the branching processes 
of the Chosen Path Tree associated with a point $x$, a pair of points $(x, y)$, and a set of points $S$.
Note that we use the same random hash functions for different points in $S$.

%\paragraph{Brute forcing.}
\subsubsection{Brute Force}
The \textsc{BruteForce} subprocedure described by Algorithm \ref{alg:bruteforce} takes two global parameters: $\mathtt{limit} \geq 1$ and $\varepsilon \geq 0$.
The parameter $\mathtt{limit}$ controls the minimum size of $S$ before we discard the \cpsj algorithm for a simple exact similarity join by brute force pairwise distance computations.
The second parameter, $\varepsilon > 0$, controls the sensitivity of the \textsc{BruteForce} step to the expected number of comparisons that a point $x \in S$ will generate if allowed to continue in the branching process.
The larger $\varepsilon$ the more aggressively we will resort to the brute force procedure.
In practice we typically think of $\varepsilon$ as a small constant, say $\varepsilon = 0.05$, 
but for some of our theoretical results we will need a sub-constant setting of $\varepsilon \approx 1/\log(n)$ to show certain running time guarantees. 
The \textsc{BruteForce} step removes a point $x$ from the Chosen Path branching process, 
instead opting to compare it against every other point $y \in S$, if it satisfies the condition
\begin{equation*}
	\frac{1}{|S| - 1}\sum_{y \in S \setminus \{ x \}}|x \cap y|/t > (1-\varepsilon)\lambda. \label{eq:bruteforce}
\end{equation*}
In the pseudocode of Algorithm \ref{alg:bruteforce} we let \texttt{count} denote a hash table that keeps track of the number of times each element $j \in [d]$ appears in $S$.
This allows us to evaluate the condition in equation \eqref{eq:bruteforce} for an element $x \in S$ in time $O(|x|)$ by rewriting it as
\begin{equation*}
\frac{1}{|S|-1}\sum_{j \in x} (\mathtt{count}[j] - 1)/t > (1-\varepsilon)\lambda.
\end{equation*}
We claim that this condition minimizes the expected number of comparisons performed by the algorithm:
Consider a node in the Chosen Path Tree associated with a set of points $S$ while running the \cpsj algorithm.
For a point ${x\in S}$, we can either remove it from $S$ immediately at a cost of $|S|-1$ comparisons, or we can choose to let continue in the branching process (possibly into several nodes) and remove it later.
The expected number of comparisons if we let it continue $k$ levels before removing it from every node that it is contained in, is given by
\begin{equation*}
	\sum_{y \in S \setminus \{ x \}} \left(\frac{1}{\lambda}\frac{|x \cap y|}{t}\right)^{k}.
\end{equation*}
This expression is convex and increasing in the similarity \mbox{$|x \cap y|/t$} between $x$ and other points $y \in S$, allowing us to state the following observation: 
\begin{observation}[Recursion]\label{obs:bruteforce}
	Let $\varepsilon = 0$ and consider a set $S$ containing a point $x \in S$ such that $x$ satisfies the recursion condition in equation~\eqref{eq:bruteforce}.
	Then the expected number of comparisons involving $x$ if we continue branching exceeds $|S|-1$ at every depth $k \geq 1$.
        If $x$ does not satisfy the condition, the opposite is observed.
\end{observation}
%
%\paragraph{Tree depth.}
\subsubsection{Tree Depth}
We proceed by bounding the maximal depth of the set of paths in the Chosen Path Tree that are explored by the \cpsj algorithm.
Having this information will allow us to bound the space usage of the algorithm and will also form part of the argument for the correctness guarantee.
Assume that the parameter \texttt{limit} in the \textsc{BruteForce} step is set to some constant value, say $\mathtt{limit} = 100$.
Consider a point $x \in S$ and let $S' = \{ y \in S \mid |x \cap y|/ t \leq (1-\varepsilon)\lambda \}$ be the subset of points in $S$ that are not too similar to $x$.
For every $y \in S'$ the expected number of vertices in the Chosen Path Tree at depth $k$ that contain both $x$ and $y$ is upper bounded by
\begin{equation*}
	\E[|F_{k}(x \cap y)|] = \left(\frac{1}{\lambda}\frac{|x \cap y|}{t}\right)^{k} \leq (1-\varepsilon)^{k} \leq e^{-\varepsilon k}.
\end{equation*}
Since $|S'| \leq n$ we use Markov's inequality to show the following bound: 
\begin{lemma}
	Let $x, y \in S$ satisfy that $|x \cap y|/ t \leq (1-\varepsilon)\lambda$ then the probability that there exists a vertex at depth $k$ 
	in the Chosen Path Tree that contains $x$ and $y$ is at most $e^{-\varepsilon k}$.
\end{lemma}
If $x$ does not share any paths with points that have similarity that falls below the threshold for brute forcing, then the only points that remain are ones that will cause $x$ to be brute forced.
This observation leads to the following probabilistic bound on the tree depth:
\begin{lemma}\label{lem:depth}
	With high probability the maximal depth of paths explored by the \cpsj algorithm is $O(\log(n) / \varepsilon)$.
\end{lemma}

%\paragraph{Correctness.}
\subsubsection{Correctness}
Let $x$ and $y$ be two sets of equal size $t$ such that $B(x, y) = |x \cap y|/t \geq \lambda$. 
We are interested in lower bounding the probability that there exists a path of length~$k$ in the Chosen Path Tree that has been chosen by both $x$ and~$y$, 
i.e. $\Pr\left[F_{k}(x \cap y)\neq\varnothing\right]$.
Agresti~\cite{agresti1974} showed an upper bound on the probability that a branching process becomes extinct after at most $k$ steps. 
We use it to show the following lower bound on the probability of a close pair of points colliding at depth $k$ in the Chosen Path Tree.
\begin{lemma}[{Agresti~\cite{agresti1974}}]
	If $\simil(x, y) \geq \lambda$ then for every $k > 0$ we have that \mbox{$\Pr[F_{k}(x \cap y) \neq \varnothing] \geq \frac{1}{k+1}$}.
\end{lemma}
The bound on the depth of the Chosen Path Tree for $x$ explored by the \cpsj algorithm in Lemma~\ref{lem:depth} then implies a lower bound on $\varphi$.

%Consider a node in the Chosen Path Tree that contains the set $S$ and assume that $x$ is not removed by the \textsc{BruteForce} step when we run \cpsj$(S, \lambda)$.
%It follows that $\sum_{y \in S} \simil(x, y) < \lambda |S| - 1$.
%This in turn allows us to bound the expected size of the set $S_j$ associated with children of the node associated with $S$.

\begin{lemma}\label{lem:correctness}
	Let $0 < \lambda < 1$ be constant. Then for every set $S$ of $|S| = n$ points the \cpsj algorithm solves the set similarity join problem with $\varphi = \Omega(\varepsilon / \log(n))$.  
\end{lemma}

\begin{remark}
  This analysis is very conservative: if either $x$ or $y$ is removed by the \textsc{BruteForce} step prior to reaching the maximum depth then it only increases the probability of collision.
  We note that similar guarantees can be obtained when using fast pseudorandom hash functions as shown in the paper introducing the \cp algorithm~\cite{christiani2017set}.
\end{remark}

%\textbf{Space usage.}
\subsubsection{Space Usage}
We can obtain a trivial bound on the space usage of the \cpsj algorithm by combining Lemma \ref{lem:depth} with the observation that every call to $\cpsj$ on the stack uses additional space at most $O(n)$. 
The result is stated in terms of working space: the total space usage when not accounting for the space required to store the data set itself 
(our algorithms use references to data points and only reads the data when performing comparisons) as well as disregarding the space used to write down the list of results. 
\begin{lemma} \label{lem:space}
	With high probability the working space of the \cpsj algorithm is at most $O(n \log (n) /\varepsilon)$.
\end{lemma}
\begin{remark}
We conjecture that the expected working space is $O(n)$ due to the size of $S$ being geometrically decreasing in expectation as we proceed down the Chosen Path Tree.
\end{remark}
%\paragraph{Running time.}
\subsubsection{Running Time}
\begin{figure}[htpb]
  \centering
  \includegraphics[width=0.45\textwidth]{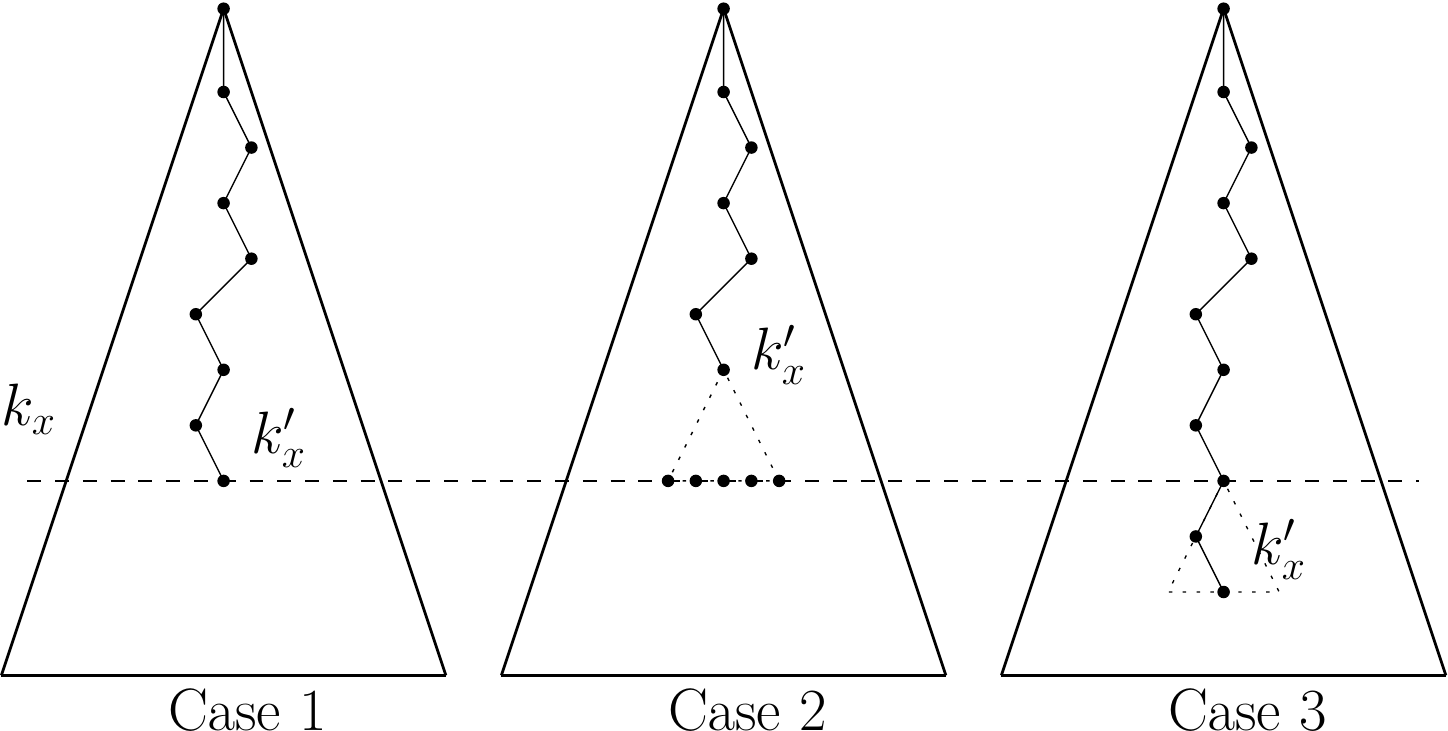}
  \caption{Path termination depth in the Chosen Path Tree}
    \label{fig:paths}
\end{figure}

We will bound the running time of a solution to the general set similarity self-join problem that uses several calls to the 
\cpsj algorithm in order to piece together a list of results $L \subseteq S \simjoin S$.
In most of the previous related work, inspired by Locality-Sensitive Hashing, the fine-grainedness of the randomized partition of space, 
here represented by the Chosen Path Tree in the \cpsj algorithm, has been controlled by a single global parameter~$k$~\cite{Gionis99, PaghSIMJOIN2015}.
In the Chosen Path setting this rule would imply that we run the splitting step without performing any brute force comparison until reaching depth $k$ 
where we proceed by comparing $x$ against every other point in nodes containing $x$, reporting close pairs. 
In recent work by Ahle et al.~\cite{ahle2017} it was shown how to obtain additional performance improvements by setting an individual depth $k_{x}$ for every $x \in S$.
We refer to these stopping strategies as global and individual, respectively.
Together with our recursion strategy, this gives rise to the following stopping criteria for when to compare a point $x$ against everything else contained in a node: 
\begin{itemize}
	\item Global: Fix a single depth $k$ for every $x \in S$.
	\item Individual: For every $x \in S$ fix a depth $k_{x}$. 
	\item Adaptive: Remove $x$ when the expected number of comparisons is non-decreasing in the tree-depth.
\end{itemize}
Let $T$ denote the running time of our similarity join algorithm. 
We aim to show the following relation between the running time between the different stopping criteria when applied to the Chosen Path Tree:
\begin{equation*}
	\E[T_{\text{Adaptive}}] \leq \E[T_{\text{Individual}}] \leq \E[T_{\text{Global}}]. 
\end{equation*}
First consider the global strategy. 
We set $k$ to balance the contribution to the running time from the expected number of vertices containing a point, given by $(1/\lambda)^{k}$,
and the expected number of comparisons between pairs of points at depth $k$, resulting in the following expected running time for the global strategy:
\begin{equation*}
	O\left(\min_{k} n (1/\lambda)^{k} +  \sum_{\substack{x,y \in S \\ x \neq y }}  (\simil(x, y) / \lambda)^{k}  \right).
\end{equation*}
The global strategy is a special case of the individual case, and it must therefore hold that $\E[T_{\text{Individual}}] \leq \E[T_{\text{Global}}]$.
The expected running time for the individual strategy is upper bounded by:
\begin{equation*}
	O\left(\sum_{x \in S}\min_{k_{x}} \left( (1/\lambda)^{k_{x}} + \sum_{y \in S \setminus \{ x \}} (\simil(x, y) / \lambda)^{k_{x}} \right) \right).
\end{equation*}
We will now argue that the expected running time of the \cpsj algorithm under the adaptive stopping criteria is no more than a constant factor greater than 
$\E[T_{\text{Individual}}]$ when we set the global parameters of the \textsc{BruteForce} subroutine as follows:
\begin{align*}
	\mathtt{limit} = \Theta(1), \\
	\varepsilon = \frac{\log(1/\lambda)}{\log n}.
\end{align*}
Let $x \in S$ and consider a path $p$ where $x$ is removed in from $S_{p}$ by the \textsc{BruteForce} step. 
Let $k_{x}'$ denote the depth of the node (length of $p$) at which $x$ is removed.
Compared to the individual strategy that removes $x$ at depth $k_{x}$ we are in one of three cases, also displayed in Figure \ref{fig:paths}.
\begin{enumerate}
	\item The point $x$ is removed from $p$ at depth $k_{x}' = k_{x}$.
	\item The point $x$ is removed from $p$ at depth $k_{x}' < k_{x}$.
	\item The point $x$ is removed from $p$ at depth $k_{x}' > k_{x}$.
\end{enumerate}
The underlying random process behind the Chosen Path Tree is not affected by our choice of termination strategy.
In the first case we therefore have that the expected running time is upper bounded by the same (conservative) expression as the one used by the individual strategy.
In the second case we remove $x$ earlier than we would have under the individual strategy. 
For every $x \in S$ we have that $k_{x} \leq 1/\varepsilon$ since for larger values of $k_{x}$ the expected number of nodes containing $x$ exceeds $n$.
We therefore have that $k_{x} - k_{x}' \leq 1/\varepsilon$.
Let $S'$ denote the set of points in the node where $x$ was removed by the \textsc{BruteForce} subprocedure.
There are two rules that could have triggered the removal of $x$: Either $|S'| = O(1)$ or the condition in equation \eqref{eq:bruteforce} was satisfied.
In the first case, the expected cost of following the individual strategy would have been $\Omega(1)$ simply from the $1/\lambda$ children containing $x$ in the next step.
This is no more than a constant factor smaller than the adaptive strategy.
In the second case, when the condition in equation \eqref{eq:bruteforce} is activated we have that the expected number of comparisons 
involving $x$ resulting from $S'$ if we had continued under the individual strategy is at least
\begin{equation*}
	(1-\varepsilon)^{1/\varepsilon}|S'| = \Omega(|S'|)
\end{equation*}
which is no better than what we get with the adaptive strategy.
In the third case where we terminate at depth $k_{x}' > k_{x}$, if we retrace the path to depth $k_{x}$ we know that $x$ was not removed in this node, 
implying that the expected number of comparisons when continuing the branching process on~$x$ is decreasing compared to removing $x$ at depth $k_{x}$.
We have shown that the expected running time of the adaptive strategy is no greater than a constant times the expected running time of the individual strategy.

We are now ready to state our main theoretical contribution, stated below as Theorem \ref{thm:main}. 
The theorem combines the above argument that compares the adaptive strategy against the individual strategy together with Lemma \ref{lem:depth} and Lemma \ref{lem:correctness}, 
and uses $O(\log^{2} n)$ runs of the \cpsj algorithm to solve the set similarity join problem for every choice of constant parameters $\lambda,\varphi$.
\begin{theorem}\label{thm:main}
	For every LSHable similarity measure and every choice of constant threshold $\lambda \in (0,1)$ and probability of recall $\varphi \in (0,1)$ 
	we can solve the $(\lambda,\varphi)$-set similarity join problem on every set $S$ of $n$ points using working space $\tilde{O}(n)$ and with expected running time
	\begin{equation*}
		\tilde{O}\left(\sum_{x \in S}\min_{k_{x}} \left( \sum_{y \in S \setminus \{ x \}} (\simil(x, y) / \lambda)^{k_{x}} + (1/\lambda)^{k_{x}} \right) \right).
	\end{equation*}
	% with working space $\tilde{O}(n)$. % High probability bound implies we can let it eat into the \varphi guarantee and just terminate if we exceed the bound
\end{theorem}

%%% Local Variables:
%%% mode: latex
%%% TeX-master: "ms"
%%% End:

%!TEX root = simfilter.tex
%-------------------------------------------
\section{Implementation} \label{sec:implementation}
%-------------------------------------------
We implement an optimized version of the \cpsj algorithm for solving the Jaccard similarity self-join problem.
In our experiments (described in Section \ref{sec:experiments}) we compare the \cpsj algorithm against 
the approximate methods of MinHash LSH~\cite{Gionis99, Broder_NETWORK97} and BayesLSH~\cite{chakrabarti2015bayesian},
as well as the AllPairs~\cite{Bayardo_WWW07} exact similarity join algorithm.
The code for our experiments is written in C\texttt{++} and uses the benchmarking framework and data sets of the recent experimental survey on exact similarity join algorithms by Mann et al.~\cite{Mann2016}.  
For our implementation we assume that each set $x$ is represented as a list of 32-bit unsigned integers. 
We proceed by describing the details of each implementation in turn. 

\subsection{Chosen Path Similarity Join} \label{sec:implementation_cpsj}
The implementation of the \cpsj algorithm follows the structure of the pseudocode in Algorithm \ref{alg:cpsjoin} and Algorithm \ref{alg:bruteforce},
but makes use of a few heuristics, primarily sampling and sketching, in order to speed things up. 
The parameter setting is discussed and investigated experimentally in section \ref{sec:parameters}.

%\textbf{Preprocessing.}
\subsubsection{Preprocessing}
Before running the algorithm we use the embedding described in section~\ref{sec:reduction}.
Specifically $t$ independent MinHash functions $h_1, \dots, h_t$ are used to map each set $x \in S$ to a list of $t$ hash values $(h_1(x), \dots, h_t(x))$.
The MinHash function is implemented using Zobrist hashing~\cite{zobrist1970new} from 32 bits to 64 bits with 8-bit characters.
We sample a MinHash function $h$ by sampling a random Zobrist hash function $g$ and let $h(x) = \arg\!\min_{j \in x} g(j)$. 
Zobrist hashing (also known as simple tabulation hashing) has been shown theoretically to have strong MinHash properties and is very fast in practice~\cite{Patrascu2012,Thorup2017}.  
We set $t = 128$ in our experiments, see discussion later. 

During preprocessing we also prepare sketches using the 1-bit minwise hashing scheme of Li and K{\"o}nig~\cite{li2011theory}. 
Let $\ell$ denote the length in 64-bit words of a sketch $\hat{x}$ of a set $x \in S$. 
We construct sketches for a data set $S$ by independently sampling $64 \times \ell$ MinHash functions $h_i$ and Zobrist hash functions $g_i$ that map from 32 bits to 1 bit. 
The $i$th bit of the sketch $\hat{x}$ is then given by $g_i(h_i(x))$.
In the experiments we set $\ell = 8$.

\medskip

%\textbf{Similarity estimation using sketches.}
\subsubsection{Similarity Estimation Using Sketches}
We use 1-bit minwise hashing sketches for fast similarity estimation in the \textsc{BruteForcePairs} and \textsc{BruteForcePoint} subroutines of the \textsc{BruteForce} step of the \cpsj algorithm.
Given two sketches, $\hat{x}$ and $\hat{y}$, we compute the number of bits in which they differ by going through the sketches word for word, computing the popcount of their XOR using the \texttt{gcc} builtin \texttt{\_mm\_popcnt\_u64} that translates into a single instruction on modern hardware. 
Let $\hat{J}(x, y)$ denote the estimated similarity of a pair of sets $(x, y)$. If $\hat{J}(x, y)$ is below a threshold $\hat{\lambda} \approx \lambda$, we exclude the pair from further consideration. If the estimated similarity is greater than $\hat{\lambda}$ we compute the exact similarity and report the pair if $J(x, y) \geq \lambda$. 

The speedup from using sketches comes at the cost of introducing false negatives:
A pair of sets $(x, y)$ with $J(x, y) \geq \lambda$ may have an estimated similarity less than $\hat{\lambda}$, causing us to miss it. 
We let $\delta$ denote a parameter for controlling the false negative probability of our sketches and set $\hat{\lambda}$ such that for sets $(x, y)$ with $J(x, y) \geq \lambda$ we have that $\Pr[\hat{J}(x, y) < \hat{\lambda}] < \delta$. 
In our experiments we set the sketch false negative probability to be $\delta = 0.05$.

\medskip

%\textbf{Splitting step.}
\subsubsection{Recursion}
In the recursive step of the \cpsj algorithm~(Algorithm \ref{alg:cpsjoin}) the set $S$ is split into buckets $S_j$ using the following heuristic: 
Instead of sampling a random hash function and evaluating it on each element $j \in x$, 
we sample an expected $1/\lambda$ elements from $[t]$ and split $S$ according to the corresponding minhash values from the preprocessing step.
This saves the linear overhead in the size of our sets $t$, reducing the time spent placing each set into buckets to $O(1)$.
Internally, a collection of sets $S$ is represented as a C\texttt{++} \texttt{std::vector<uint32\_t>} of set ids.
%The collection of buckets $S_j$ is implemented using Google's \texttt{dense\_hash} hash map implementation from the \texttt{sparse\_hash} package~\cite{sparsehash}.

\medskip

%\textbf{BruteForce step.}
\subsubsection{BruteForce}
Having reduced the overhead for each set $x \in S$ to $O(1)$ in the splitting step, we wish to do the same for the \textsc{BruteForce} step (described in Algorithm \ref{alg:bruteforce}), 
at least in the case where we do not call the \textsc{BruteForcePairs} or \textsc{BruteForcePoint} subroutines.
The main problem is that we spend time $O(t)$ for each set when constructing the \texttt{count} hash map and estimating the average similarity of $x$ to sets in $S \setminus \{x\}$.
To get around this we construct a 1-bit minwise hashing sketch $\hat{s}$ of length $64 \times \ell$ for the set $S$ using sampling and our precomputed 1-bit minwise hashing sketches.
The sketch $\hat{s}$ is constructed as follows: Randomly sample $64 \times \ell$ elements of $S$ and set the $i$th bit of $\hat{s}$ to be the $i$th bit of the $i$th sample from $S$.
This allows us to estimate the average similarity of a set $x$ to sets in $S$ in time $O(\ell)$ using word-level parallelism. 
A set $x$ is removed from $S$ if its estimated average similarity is greater than $(1 - \varepsilon)\lambda$. 
To further speed up the running time we only call the \textsc{BruteForce} subroutine once for each call to \cpsj, 
calling \textsc{BruteForcePoint} on all points that pass the check rather than recomputing $\hat{s}$ each time a point is removed.
Pairs of sets that pass the sketching check are verified using the same verification procedure as the \all implementation by Mann et al.~\cite{Mann2016}.
In our experiments we set the parameter $\varepsilon = 0.1$.
Duplicates are removed by sorting and performing a single linear scan.

\subsubsection{Repetitions}
In theory, for any constant desired recall $\varphi \in (0,1)$ it suffices with $O(\log^2 n)$ independent repetitions of the \cpsj algorithm. 
In practice this number of repetitions is prohibitively large and we therefore set the number of independent repetitions used in our experiments to be fixed at ten. 
With this setting we were able to achieve more than $90\%$ recall across all datasets and similarity thresholds.

\subsection{MinHash LSH} \label{sec:minhash}
We implement a locality-sensitive hashing similarity join using MinHash according to the pseudocode in Algorithm~\ref{alg:minhash}.
A single run of the \mh algorithm can be divided into two steps: 
First we split the sets into buckets according to the hash values of $k$ concatenated MinHash functions $h(x) = (h_1(x), \dots, h_k(x))$.
Next we iterate over all non-empty buckets and run \textsc{BruteForcePairs} to report all pairs of points with similarity above the threshold $\lambda$.
The \textsc{BruteForcePairs} subroutine is shared between the \mh and \cpsj implementation.
\mh therefore uses 1-bit minwise sketches for similarity estimation in the same way as in the implementation of the \cpsj algorithm described above. 

The parameter $k$ can be set for each dataset and similarity threshold $\lambda$ to minimize the combined cost of lookups and similarity estimations performed by algorithm.
This approach was mentioned by Cohen et al.~\cite{cohen2001finding} but we were unable to find an existing implementation.
In practice we set $k$ to the value that results in the minimum estimated running time when running the first part (splitting step) of the algorithm for values of $k$ in the range $\{2, 3, \dots, 10\}$ and estimating the running time by looking at the number of buckets and their sizes. 
Once $k$ is fixed we know that each repetition of the algorithm has probability at least $\lambda^k$ of reporting a pair $(x, y)$ with $J(x, y) \geq \lambda$. 
For a desired recall $\varphi$ we can therefore set $L = \lceil \ln(1/(1-\varphi)) / \lambda^k \rceil$.
In our experiments we report the actual number of repetitions required to obtain a desired recall rather than using the setting of $L$ required for worst-case guarantees.
\begin{algorithm}
\SetKwInOut{Params}{Parameters}
\SetKwArray{Buckets}{buckets}
\DontPrintSemicolon
\Params{$k \geq 1, L \geq 1$.}
\For{$i \leftarrow 1$ \KwTo $L$} {
\emph{Initialize hash map \Buckets{\,}.}\;
\emph{Sample $k$ MinHash fcts.} $h \leftarrow (h_1, \dots, h_k)$\;
\For{$x \in S$} {
	\Buckets{$h(x)$} $\leftarrow$ \Buckets{$h(x)$} $\cup$ $\{x\}$\; 
}
\For{$S' \in$ \Buckets} {
	$\textsc{BruteForcePairs}(S', \lambda)$\;
}
}
\caption{\textsc{MinHash}$(S, \lambda)$} \label{alg:minhash}
\end{algorithm}

\subsection{AllPairs}
To compare our approximate methods against a state-of-the-art exact similarity join we use Bayardo et al.'s \all algorithm~\cite{Bayardo_WWW07} as recently implemented in the set similarity join study by Mann et al.~\cite{Mann2016}. 
The study by Mann et al. compares implementations of several different exact similarity join methods and finds that the simple \all algorithm is most often the fastest choice. 
Furthermore, for Jaccard similarity, the \all algorithm was at most $2.16$ times slower than the best out of six different competing algorithm across all the data sets and similarity thresholds used, 
and for most runs \all is at most $11\%$ slower than the best exact algorithm (see Table 7 in Mann et al.~\cite{Mann2016}). 
Since our experiments run in the same framework and using the same datasets and with the same thresholds as Mann et al.'s study, 
we consider their \all implementation to be a good representative of exact similarity join methods for Jaccard similarity.  

\subsection{BayesLSH}
For a comparison against previous experimental work on approximate similarity joins we use an implementation of \blsh in C as provided by the \blsh authors~\cite{chakrabarti2015bayesian, bayeslsh}.
The BayesLSH package features a choice between \all and LSH as candidate generation method. 
For the verification step there is a choice between \blsh and \blsh-lite.
Both verification methods use sketching to estimate similarities between candidate pairs.
The difference between BayesLSH and BayesLSH-lite is that the former uses sketching to estimate the similarity of pairs that pass the sketching check, 
whereas the latter uses an exact similarity computation if a pair passes the sketching check.
Since the approximate methods in our \cpsj and \mh implementations correspond to the approach of BayesLSH-lite we restrict our experiments to this choice of verification algorithm.
In our experiments we will use \blsh to represent the fastest of the two candidate generation methods, combined with BayesLSH-lite for the verification step.

%%% Local Variables:
%%% mode: latex
%%% TeX-master: "ms"
%%% End:

%!TEX root = simfilter.tex
%-------------------------------------------
\section{Experiments} \label{sec:experiments}
%-------------------------------------------
We run experiments using the implementations of \cpsj, \mh, \blsh, and \all described in the previous section.
In the experiments we perform self-joins under Jaccard similarity for similarity thresholds $\lambda \in \{0.5, 0.6, 0.7, 0.8, 0.9 \}$.
We are primarily interested in measuring the join time of the algorithms, but we also look at the number of candidate pairs $(x,y)$ considered by the algorithms during the join as a measure of performance.  
Note that the preprocessing step of the approximate methods only has to be performed once for each set and similarity measure, 
and can be re-used for different similarity joins, we therefore do not count it towards our reported join times.
In practice the preprocessing time is at most a few minutes for the largest data sets.

%\paragraph{Data sets.}
\subsubsection{Data Sets}
The performance is measured across $10$ real~world data sets along with $4$ synthetic data sets described in Table \ref{tab:datasets}. 
All datasets except for the TOKENS datasets were provided by the authors of~\cite{Mann2016} where descriptions and sources for each data set can also be found. 
Note that we have excluded a synthetic ZIPF dataset used in the study by Mann et al.\cite{Mann2016} due to it having no results for our similarity thresholds of interest.
Experiments are run on versions of the datasets where duplicate records are removed and any records containing only a single token are ignored.
\begin{table}
	\centering
	\caption{Dataset size, average set size, and average number of sets that a token is contained in.}
	\label{tab:datasets}
	\small

	\begin{tabular}{lrrr} \toprule
		Dataset & \# sets / $10^6$ & avg. set size & sets / tokens \\\midrule
		AOL        & $7.35$ &   $3.8$ & $18.9$ \\
		BMS-POS    & $0.32$ &   $9.3$ & $1797.9$ \\
		DBLP       & $0.10$ &  $82.7$ & $1204.4$ \\
		ENRON      & $0.25$ & $135.3$ & $29.8$ \\
		FLICKR     & $1.14$ &  $10.8$ & $16.3$ \\
		LIVEJ      & $0.30$ &  $37.5$ & $15.0$ \\
		KOSARAK    & $0.59$ &  $12.2$ & $176.3$ \\
		NETFLIX    & $0.48$ & $209.8$ & $5654.4$ \\
		ORKUT      & $2.68$ & $122.2$ & $37.5$ \\
		SPOTIFY    & $0.36$ &  $15.3$ & $7.4$ \\
		UNIFORM    & $0.10$ &  $10.0$ & $4783.7$ \\
		%ZIPF       & $0.10$ &  $50.0$ & $49.1$ \\
		TOKENS10K  & $0.03$ & $339.4$ & $10000.0$ \\
		TOKENS15K  & $0.04$ & $337.5$ & $15000.0$ \\
		TOKENS20K  & $0.06$ & $335.7$ & $20000.0$ \\ \bottomrule
	\end{tabular}

\end{table}

In addition to the datasets from the study of Mann et al. we add three synthetic datasets TOKENS10K, TOKENS15K, and TOKENS20K, designed to showcase the robustness of the approximate methods.
These datasets have relatively few unique tokens, but each token appears in many sets. 
Each of the TOKENS datasets were generated from a universe of $1000$ tokens ($d = 1000$) and each token is contained in respectively, $10,000$, $15,000$, and $20,000$ different sets as denoted by the name.
The sets in the TOKENS datasets were generated by sampling a random subset of the set of possible tokens,
rejecting tokens that had already been used in more than the maximum number of sets ($10,000$ for TOKENS10K).
To sample sets with expected Jaccard similarity $\lambda'$ the size of our sampled sets should be set to $(2\lambda'/(1+\lambda'))d$. 
For $\lambda' \in \{0.95, 0.85, 0.75, 0.65, 0.55\}$ the TOKENS datasets each have $100$ random sets planted with expected Jaccard similarity $\lambda'$.
This ensures an increasing number of results for our experiments where we use thresholds $\lambda \in \{0.5, 0.6, 0.7, 0.8, 0.9 \}$. 
The remaining sets have expected Jaccard similarity $0.2$.
We believe that the TOKENS datasets give a good indication of the performance on real-world data that has the property that most tokens appear in a large number of sets.

%\paragraph{Recall.}
\subsubsection{Recall}
In our experiments we aim for a recall of at least $90\%$ for the approximate methods. 
In order to achieve this for the \cpsj and \mh algorithms we perform a number of repetitions after the preprocessing step, stopping when the desired recall has been achieved.
This is done by measuring the recall against the recall of \all and stopping when reaching $90\%$.
In practice this approach is not feasible as the size of the true result set is not known.
However, it can be efficiently estimated using sampling if it is not too small.
Another approach is to perform the number of repetitions required to obtain the theoretical guarantees on recall as described for \cpsj in Section \ref{sec:analysis} and for \mh in Section \ref{sec:minhash}.
Unfortunately, with our current analysis of the \cpsj algorithm the number of repetitions required to guarantee theoretically a recall of $90\%$ far exceeds the number required in practice as observed in our experiments where ten independent repetitions always suffice. 
For \blsh using LSH as the candidate generation method, the recall probability with the default parameter setting is $95\%$, although we experience a recall closer to $90\%$ in our experiments.

%\paragraph{Hardware.}
\subsubsection{Hardware}
All experiments were run on an Intel Xeon E5-2690v4 CPU at 2.60GHz with $35$MB L$3$,$256$kB L$2$ and $32$kB L$1$ cache and $512$GB of RAM.
Since a single experiment is always confined to a single CPU core we ran several experiments in parallel~\cite{Tange2011a} to better utilize our hardware.
%-------------------------------------------
\subsection{Results}
%-------------------------------------------
%\paragraph{Join time.}
\subsubsection{Join Time}
%-------------------------------------------
Table \ref{tab:jointimes} shows the average join time in seconds over five independent runs, when approximate methods are required to have at least $90\%$ recall.
We have omitted timings for \blsh since it was always slower than all other methods, and in most cases it timed out after 20 minutes when using LSH as candidate generation method.
The join time for \mh is always greater than the corresponding join time for \cpsj except in a single setting: the dataset KOSARAK with threshold $\lambda = 0.5$.
Since \cpsj is typically $2-4\times$ faster than \mh we can restrict our attention to comparing \all and \cpsj where the picture becomes more interesting.

\begin{table*}
	\caption{Join time in seconds for \cpsj (CP), \mh (MH) and \all (ALL) with at least $\ge90\%$ recall.}
	\label{tab:jointimes}
\resizebox{\textwidth}{!}{%	
	\footnotesize
\begin{filecontents*}{results_table.csv}
File,AA,ACP,AMH,BA,BCP,BMH,CA,CCP,CMH,DA,DCP,DMH,EA,ECP,EMH
AOL, 483.5,\bf{ 362.1},1329.9, 117.8,\bf{ 113.4}, 444.2,\bf{  13.7},  42.2, 152.9,\bf{   4.2},  34.6, 100.6,\bf{   1.6},  21.0,  43.8
BMS-POS,  62.5,\bf{  27.0},  40.0,  20.9,\bf{   7.1},  13.7,   5.6,\bf{   2.7},   5.6,\bf{   1.3},   2.0,   3.9,\bf{   0.2},   0.9,   1.4
DBLP, 127.9,\bf{   9.2},  22.1,  63.8,\bf{   2.5},  10.1,  27.4,\bf{   1.1},   3.7,   7.8,\bf{   0.6},   1.8,   0.8,\bf{   0.3},   0.7
ENRON,  78.0,\bf{   6.9},  16.4,  23.2,\bf{   4.4},   9.9,   6.0,\bf{   2.4},   6.3,\bf{   1.6},   1.6,   2.7,\bf{   0.4},   0.7,   1.7
FLICKR,\bf{  17.2},  48.6,  68.0,\bf{   6.0},  30.9,  37.2,\bf{   2.5},  13.8,  21.3,\bf{   1.0},   6.3,  11.3,\bf{   0.3},   3.4,   5.2
KOSARAK,\bf{  73.1}, 377.9, 311.1,\bf{  14.4},  62.7,  89.2,\bf{   1.6},   7.2,  16.1,\bf{   0.5},   3.9,   9.9,\bf{   0.1},   1.2,   2.6
LIVEJ, 571.7,\bf{ 131.3}, 279.4, 145.3,\bf{  48.7}, 129.6,  30.6,\bf{  28.2},  52.9,\bf{   7.1},  16.2,  41.0,\bf{   1.5},   9.2,  12.6
NETFLIX,1354.7,\bf{  25.3}, 121.8, 520.4,\bf{   8.2},  60.0, 177.3,\bf{   4.8},  22.6,  46.2,\bf{   2.4},  14.1,   5.4,\bf{   1.6},   5.8
ORKUT, 359.7,\bf{  26.5}, 115.7, 106.4,\bf{  15.4},  60.1,  36.3,\bf{   8.0},  25.1,  12.2,\bf{   7.4},  19.7,\bf{   3.7},   4.8,  13.3
SPOTIFY,\bf{   0.5},   2.5,   9.3,\bf{   0.3},   1.5,   3.4,\bf{   0.2},   1.0,   2.6,\bf{   0.1},   1.0,   1.9,\bf{   0.1},   0.5,   0.6
TOKENS10K, 312.1,\bf{   3.4},   4.8, 236.8,\bf{   2.9},   3.9, 164.0,\bf{   1.5},   1.7, 114.9,\bf{   0.6},   1.2,  63.2,\bf{   0.2},   0.4
TOKENS15K, 688.4,\bf{   4.4},   6.2, 535.3,\bf{   4.0},   7.1, 390.4,\bf{   1.8},   3.7, 258.2,\bf{   0.7},   1.7, 140.0,\bf{   0.2},   0.7
TOKENS20K,1264.1,\bf{   5.7},  12.0, 927.0,\bf{   4.0},  11.4, 698.4,\bf{   2.1},   4.5, 494.3,\bf{   0.8},   2.2, 273.4,\bf{   0.3},   0.8
UNIFORM005,  54.1,\bf{   3.9},   6.6,  27.6,\bf{   1.6},   3.0,  10.5,\bf{   0.9},   1.4,   3.6,\bf{   0.5},   1.0,   0.4,\bf{   0.1},   0.3
\end{filecontents*}
\csvreader[tabular=l|rrr|rrr|rrr|rrr|rrr
, table head=\toprule&\multicolumn{3}{c}{Threshold $0.5$}&\multicolumn{3}{c}{Threshold $0.6$}&\multicolumn{3}{c}{Threshold $0.7$}&\multicolumn{3}{c}{Threshold $0.8$}&\multicolumn{3}{c}{Threshold $0.9$}\\
	\midrule Dataset & CP & MH & ALL & CP & MH & ALL & CP & MH & ALL & CP & MH & ALL & CP & MH & ALL \\ \midrule
, table foot=, head to column names
, late after line=\\
, late after last line =\\\bottomrule]{results_table.csv}{}
{\File& \ACP & \AMH & \AA & \BCP & \BMH & \BA & \CCP & \CMH & \CA &\DCP & \DMH & \DA &\ECP & \EMH & \EA}
}
\end{table*}

\begin{figure}
  \centering
  \includegraphics[width=0.5\textwidth]{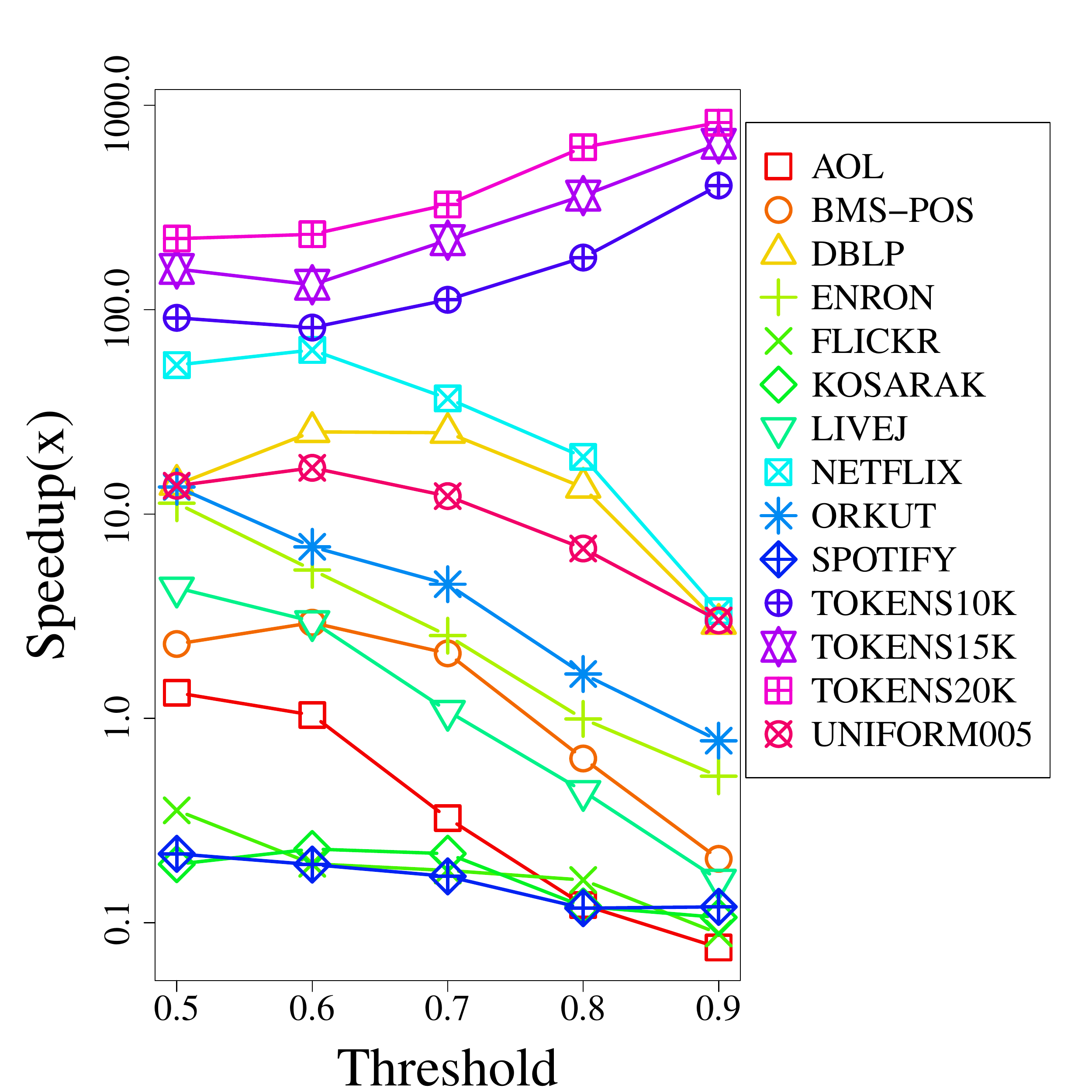}
  \caption{Join time of \cpsj with at least $90\%$ recall relative to \all.}
  \label{fig:speed}
\end{figure}

Figure \ref{fig:speed} shows the join time speedup that \cpsj achieves over \all. 
We achieve speedups of between $2-50\times$ for most of the datasets, with greater speedups at low similarity thresholds.
For a number of the datasets the \cpsj algorithm is slower than \all for the thresholds considered here.
Comparing with Table \ref{tab:datasets} it seems that \cpsj generally performs well on most data sets where tokens are contained in a large number of sets on average (NETFLIX, UNIFORM, DBLP),
but is beaten by \all on datasets that have a lot of ``rare'' tokens (SPOTIFY, FLICKR, AOL).
This is logical because rare tokens are exploited by the sorted prefix-filtering in \all.
Without rare tokens \all will be reading long inverted indexes.
This is a common theme among all the current state-of-the-art exact methods examined in \cite{Mann2016}, including \pp.
\cpsj is robust in the sense that it does not depend on the presence of rare tokens in the data to perform well.
This difference is showcased with the synthetic TOKEN data sets.

%\paragraph{BayesLSH.}
\subsubsection{BayesLSH}
The poor performance of \blsh compared to the other algorithms (\blsh was always slower) can most likely be tracked down to differences in the implementation of the candidate generation methods of \blsh.
The \blsh implementation uses an older implementation of \all compared to the implementation by Mann et al.~\cite{Mann2016} which was shown to yield performance improvements by using a more efficient verification procedure. 
The LSH candidate generation method used by \blsh corresponds to the \mh splitting step, but with $k$ (the number of hash functions) fixed to one. 
Our technique for choosing $k$ in the \mh algorithm, aimed at minimizing the total join time, typically selects $k \in \{3,4,5,6\}$ in the experiments. 
It is therefore likely that \blsh can be competitive with the other techniques by combining it with other candidate generation procedures.
Further experiments to compare the performance of BayesLSH sketching to 1-bit minwise sketching for different parameter settings and similarity thresholds would also be instructive.

%\paragraph{TOKEN datasets.}
\subsubsection{TOKEN datasets}
The TOKENS datasets clearly favor the approximate join algorithms where \cpsj is two to three orders of magnitude faster than \all.
By increasing the number of times each token appears in a set we can make the speedup of \cpsj compared to \all arbitrarily large as shown by the progression from TOKENS10 to TOKENS20.
The \all algorithm generates candidates by searching through the lists of sets that contain a particular token, starting with rare tokens.
Since every token appears in a large number of sets every list will be long.

Interestingly, the speedup of \cpsj is even greater for higher similarity thresholds.
We believe that this is due to an increase in the gap between the similarity of sets to be reported and the remaining sets that have an average Jaccard similarity of $0.2$.
This is in line with our theoretical analysis of \cpsj and most theoretical work on approximate similarity search 
where the running time guarantees usually depend on the approximation factor.

%-------------------------------------------
%\paragraph{Candidates and verification.}
\subsubsection{Candidates and Verification}
%-------------------------------------------
Table \ref{tab:candidates} compares the number of pre-candidates, candidates, and results generated by the \all and \cpsj algorithms where the desired recall for \cpsj is set to be greater than $90\%$.
For \all the number of pre-candidates denotes all pairs $(x, y)$ investigated by the algorithm that pass checks on their size so that it is possible that $J(x, y) \geq \lambda$.
The number of candidates is simply the number of unique pre-candidates as duplicate pairs are removed explicitly by the \all algorithm.

For \cpsj we define the number of pre-candidates to be all pairs $(x, y)$ considered by the \textsc{BruteForcePairs} and \textsc{BruteForcePoint} subroutines of Algorithm \ref{alg:bruteforce}.
The number of candidates are pre-candidate pairs that pass size checks (similar to \all) and the 1-bit minwise sketching check as described in Section \ref{sec:implementation_cpsj}.
Note that for \cpsj the number of candidates may still contain duplicates as this is inherent to the approximate method for candidate generation. 
Removing duplicates though the use of a hash table would drastically increase the space usage of \cpsj.
For both \all and \cpsj the number of candidates denotes the number of points that are passed to the exact similarity verification step of the \all implementation of Mann et al.~\cite{Mann2016}.

Table \ref{tab:candidates} shows that for \all there is not a great difference between the number of pre-candidates and number of candidates, 
while for \cpsj the number of candidates is usually reduced by one or two orders of magnitude for datasets where \cpsj performs well.
For datasets where \cpsj performs poorly such as AOL, FLICKR, and KOSARAK there is less of a decrease when going from pre-candidates to candidates.
It would appear that this is due to many duplicate pairs from the candidate generation step and not a failure of the sketching technique.
\begin{figure*}[tb]
\subfloat[\texttt{limit} $\in \{10, 50, 100, 250, 500 \}$]{\includegraphics[width = 0.7\columnwidth]{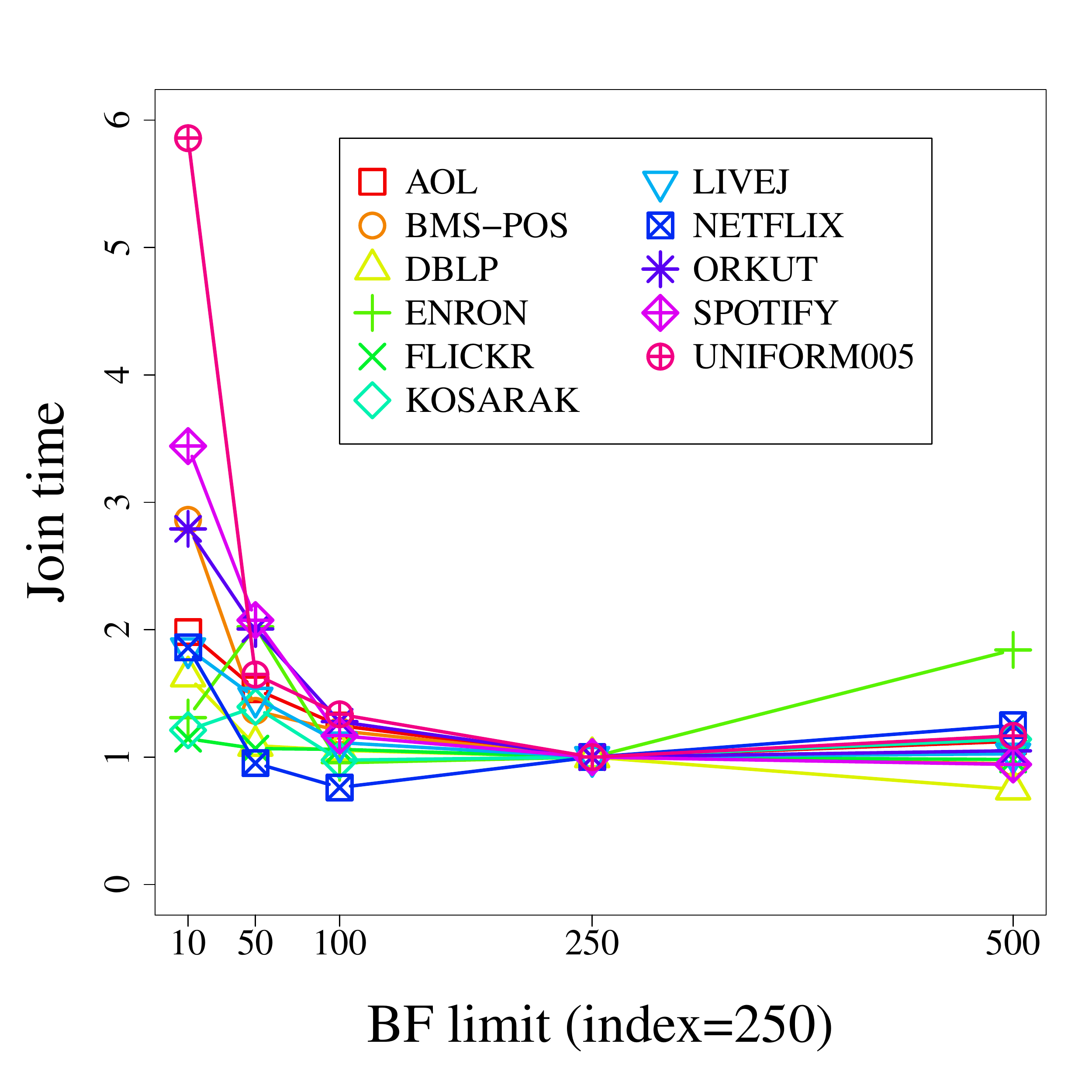}} 
\subfloat[$\varepsilon \in \{0.0, 0.1, 0.2, 0.3, 0.4, 0.5 \}$]{\includegraphics[width = 0.7\columnwidth]{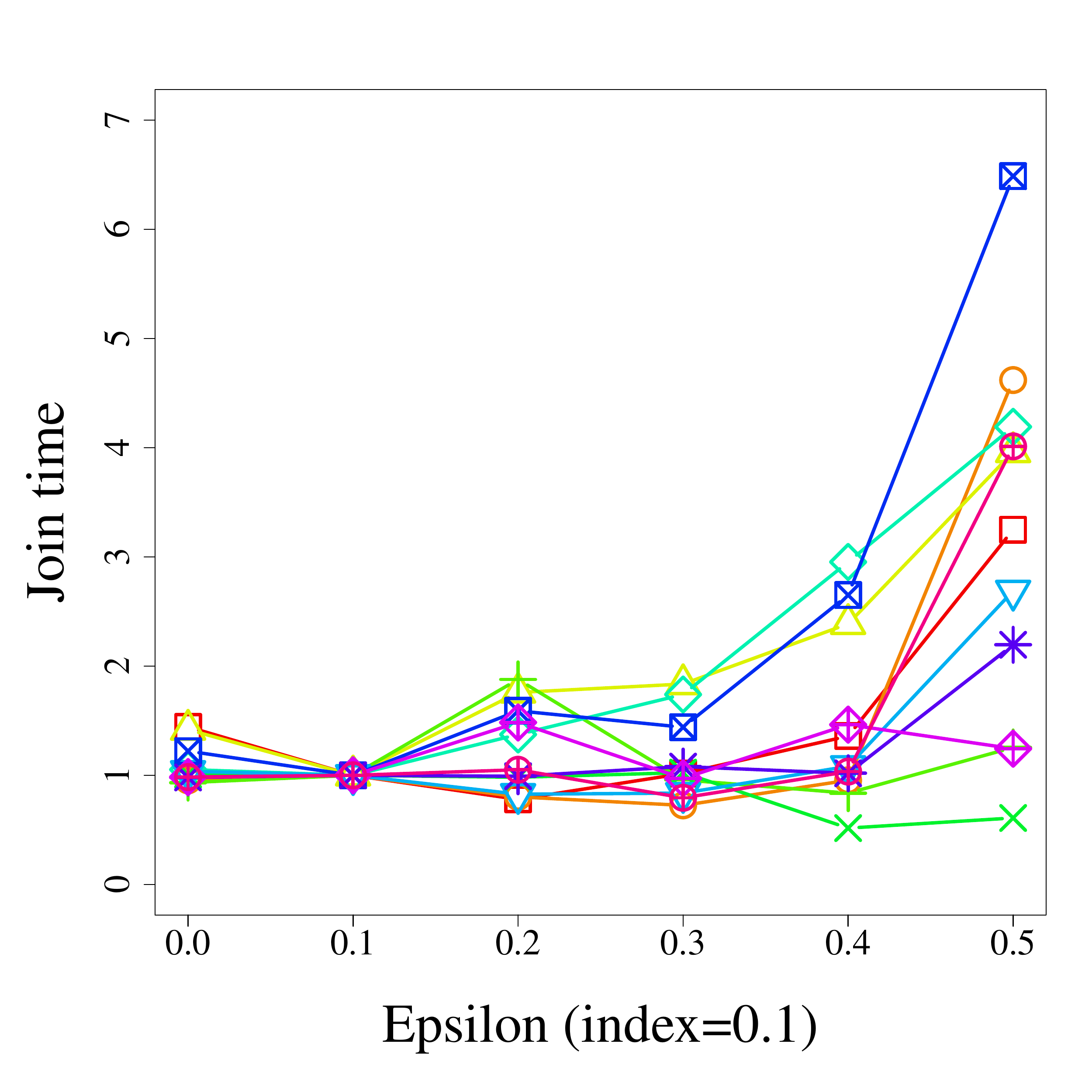}}    
\subfloat[$w \in \{1, 2, 4, 8, 16 \}$]{\includegraphics[width = 0.7\columnwidth]{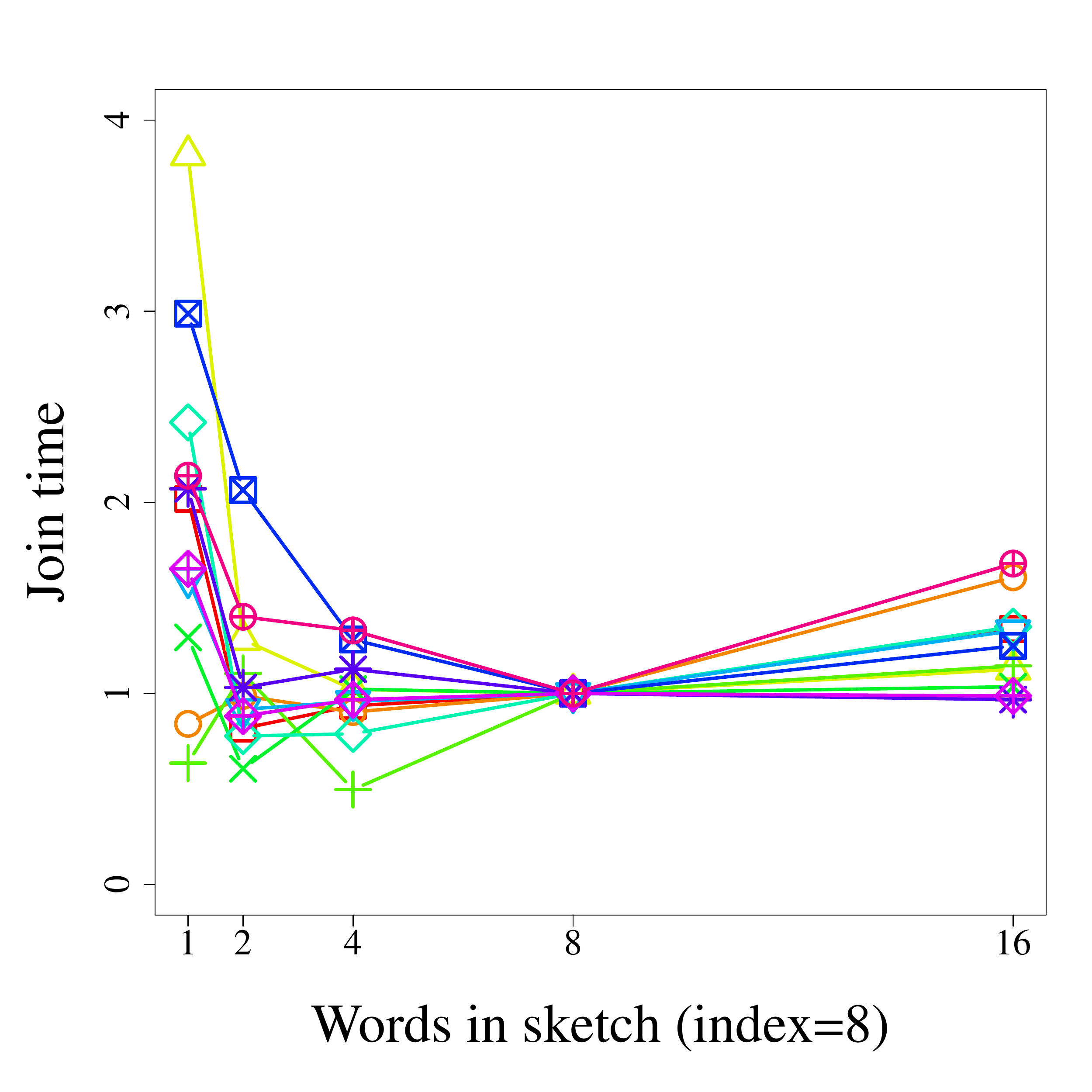}}    
\caption{Relative join time for \cpsj with at least $80\%$ recall and similarity threshold $\lambda = 0.5$ for different parameter settings of \texttt{limit}, $\varepsilon$, and $w$.} 
\label{fig:parameters}
\end{figure*}
%-------------------------------------------
\subsection{Parameters} \label{sec:parameters}
%-------------------------------------------
To investigate how parameter settings affect the performance of the \cpsj algorithm we run experiments where we vary the brute force parameter \texttt{limit},
the brute force aggressiveness parameter $\varepsilon$, and the sketch length in words $\ell$.
Table \ref{tab:parameters} shows the parameter settings used during theses experiments and the final setting used for our join time experiments.
\begin{table}[b]
\caption{Parameters of the \cpsj algorithm, their setting during parameter experiments, and their setting for the final join time experiments}
\label{tab:parameters}
\small
\centering
	\begin{tabular}{llrr} \toprule
		Parameter & Description & Test & Final \\ \midrule
		\texttt{limit} & Brute force limit & $100$ & $250$ \\
		$\ell$ & Sketch word length & $4$ & $8$ \\
		$t$ & Size of MinHash set & $128$ & $128$ \\
		$\varepsilon$ & Brute force aggressiveness & $0.0$ & $0.1$ \\ 
		$\delta$ & Sketch false negative prob. & $0.1$ & $0.05$ \\ \bottomrule 
	\end{tabular}
\end{table}
Figure \ref{fig:parameters} shows the \cpsj join time for different settings of the parameters.
By picking one parameter at a time we are obviously ignoring possible interactions between the parameters, but the stability of the join times lead us to believe that these interactions have limited effect.

Figure \ref{fig:parameters} (a) shows the effect of the brute force limit on the join time. 
Lowering \texttt{limit} causes the join time to increase due to a combination of spending more time splitting sets into buckets and due to the lower probability of recall from splitting at a deeper level. The join time is relatively stable for \texttt{limit} $\in \{100, 250, 500\}$.  

Figure \ref{fig:parameters} (b) shows the effect of brute force aggressiveness on the join time. 
As we increase $\varepsilon$, sets that are close to the other elements in their buckets are more likely to be removed by brute force comparing them to all other points.
The tradeoff here is between the loss of probability of recall by letting a point continue in the $\cp$ branching process versus the cost of brute forcing the point.
The join time is generally increasing with $\varepsilon$, but it turns out that $\varepsilon = 0.1$ is a slightly better setting than $\varepsilon = 0.0$ for almost all data sets.

Figure \ref{fig:parameters} (c) shows the effect of sketch length on the join time.
There is a trade-off between the sketch similarity estimation time and the precision of the estimate, leading to fewer false positives.
For a similarity threshold of $\lambda = 0.5$ using only a single word negatively impacts the performance on most datasets compared to using two or more words.
The cost of using longer sketches seems neglible as it is only a few extra instructions per similarity estimation so we opted to use $\ell = 8$ words in our sketches.
\begin{table}[t]
\caption{Number of pre-candidates, candidates and results for ALL and CP with at least $90\%$ recall.}
\label{tab:candidates}
\footnotesize
\centering
\renewcommand*{\arraystretch}{1.05}
\begin{tabular}{l|rr|rr}
\toprule
Dataset    &   \multicolumn{2}{c}{Threshold $0.5$}&\multicolumn{2}{c}{Threshold $0.7$} \\ 
		   &\multicolumn{1}{c}{ALL}&\multicolumn{1}{c}{CP} &	\multicolumn{1}{c}{ALL}&\multicolumn{1}{c}{CP} 	  \\ \midrule
           & 8.5E+09 & 7.4E+09 & 6.2E+08 & 2.9E+09 \\
AOL        & 8.5E+09 & 1.4E+09 & 6.2E+08 & 3.1E+07 \\
           & 1.3E+08 & 1.2E+08 & 1.6E+06 & 1.5E+06 \\ \hline
           & 2.0E+09 & 9.2E+08 & 2.7E+08 & 3.3E+08 \\
BMS-POS    & 1.8E+09 & 1.7E+08 & 2.6E+08 & 4.9E+06 \\
           & 1.1E+07 & 1.0E+07 & 2.0E+05 & 1.8E+05 \\ \hline
           & 6.6E+09 & 4.6E+08 & 1.2E+09 & 1.3E+08 \\
DBLP       & 1.9E+09 & 4.6E+07 & 7.2E+08 & 4.3E+05 \\
           & 1.7E+06 & 1.6E+06 & 9.1E+03 & 8.5E+03 \\ \hline
           & 2.8E+09 & 3.7E+08 & 2.0E+08 & 1.5E+08 \\
ENRON      & 1.8E+09 & 6.7E+07 & 1.3E+08 & 2.1E+07 \\
           & 3.1E+06 & 2.9E+06 & 1.2E+06 & 1.2E+06 \\ \hline
           & 5.7E+08 & 2.1E+09 & 9.3E+07 & 9.0E+08 \\
FLICKR     & 4.1E+08 & 1.1E+09 & 6.3E+07 & 3.8E+08 \\
           & 6.6E+07 & 6.1E+07 & 2.5E+07 & 2.3E+07 \\ \hline
           & 2.6E+09 & 4.7E+09 & 7.4E+07 & 4.2E+08 \\
KOSARAK    & 2.5E+09 & 2.1E+09 & 6.8E+07 & 2.1E+07 \\
           & 2.3E+08 & 2.1E+08 & 4.4E+05 & 4.1E+05 \\  \hline
           & 9.0E+09 & 2.8E+09 & 5.8E+08 & 1.2E+09 \\
LIVEJ      & 8.3E+09 & 3.6E+08 & 5.6E+08 & 1.8E+07 \\
           & 2.4E+07 & 2.2E+07 & 8.1E+05 & 7.6E+05 \\ \hline
           & 8.6E+10 & 1.3E+09 & 1.0E+10 & 4.3E+08 \\
NETFLIX    & 1.3E+10 & 3.1E+07 & 3.4E+09 & 6.4E+05 \\
           & 1.0E+06 & 9.5E+05 & 2.4E+04 & 2.2E+04 \\ \hline
           & 5.1E+09 & 1.1E+09 & 3.0E+08 & 7.2E+08 \\
ORKUT      & 3.9E+09 & 1.3E+06 & 2.6E+08 & 8.1E+04 \\
           & 9.0E+04 & 8.4E+04 & 5.6E+03 & 5.3E+03 \\ \hline
           & 5.0E+06 & 1.2E+08 & 4.7E+05 & 8.5E+07 \\
SPOTIFY    & 4.8E+06 & 3.1E+05 & 4.6E+05 & 2.7E+03 \\
           & 2.0E+04 & 1.8E+04 & 2.0E+02 & 1.9E+02 \\ \hline
           & 1.5E+10 & 1.7E+08 & 8.1E+09 & 4.9E+07 \\
TOKENS10K  & 4.1E+08 & 5.7E+06 & 4.1E+08 & 1.9E+06 \\
           & 1.3E+05 & 1.3E+05 & 7.4E+04 & 6.9E+04 \\ \hline
           & 3.6E+10 & 3.0E+08 & 1.9E+10 & 8.1E+07 \\
TOKENS15K  & 9.6E+08 & 7.2E+06 & 9.6E+08 & 1.9E+06 \\
           & 1.4E+05 & 1.3E+05 & 7.5E+04 & 6.9E+04 \\ \hline
           & 6.4E+10 & 4.4E+08 & 3.4E+10 & 1.0E+08 \\
TOKENS20K  & 1.7E+09 & 8.8E+06 & 1.7E+09 & 1.9E+06 \\
           & 1.4E+05 & 1.4E+05 & 7.9E+04 & 7.4E+04 \\ \hline
           & 2.5E+09 & 3.7E+08 & 6.5E+08 & 1.3E+08 \\
UNIFORM005 & 2.0E+09 & 9.5E+06 & 6.1E+08 & 3.9E+04 \\
           & 2.6E+05 & 2.4E+05 & 1.4E+03 & 1.3E+03 \\ \bottomrule 
\end{tabular}
\end{table}

%%% Local Variables:
%%% mode: latex
%%% TeX-master: "paper"
%%% End:

\section{Conclusion}
We provided experimental and theoretical results on a new randomized set similarity join algorithm, \cpsj, and compared it experimentally to state-of-the-art exact and approximate set similarity join algorithms.
\cpsj is typically $2-4$ times faster than previous approximate methods.
Compared to exact methods it obtains speedups of more than an order of magnitude on real-world datasets, while keeping the recall above $90\%$.
Among the datasets used in these experiments we note that NETFLIX and FLICKR represents two archetypes.
On average a token in the NETFLIX dataset appears in more than $5000$ sets while on average a token in the FLICKR dataset appears in less than $20$ sets. 
Our experiments indicate that \cpsj brings large speedups to the NETFLIX type datasets, while it is hard to improve upon the perfomance of \all on the FLICKR type.

A direction for future work could be to tighten and simplify the theoretical analysis.
We conjecture that the running time of the algorithm can be bounded by a simpler function of the sum of similarities between pairs of points in $S$.
We note that recursive methods such as ours lend themselves well to parallel and distributed implementations since most of the computation happens in independent, recursive calls. Further investigating this is an interesting possibility.

\clearpage
\section*{Acknowledgments}
The authors would like to thank Willi Mann for making the source code and data sets of~\cite{Mann2016} available, Aniket Chakrabarti for information about the implementation of BayesLSH, and the anonymous reviewers for useful suggestions.

\bibliographystyle{IEEEtran}
\bibliography{IEEEabrv,biblio}
\end{document}